\documentclass[aps,prl,showpacs, twocolumn]{revtex4-1}

\usepackage[latin2]{inputenc}
\usepackage{graphicx}
\usepackage{bm}
\usepackage{color}
\begin{document}

\title{Observation of Ion-wave Satellites to Laser Harmonics in Intense Picosecond Laser-Solid Interaction}

\author{R.S. Marjoribanks}
  \email{robin.marjoribanks@utoronto.ca}
\author{L. Zhao}
\author{F.W. Budnik}
\author{G. Kulcs\'ar}
\affiliation{
Department of Physics, 
University of Toronto\\
60 Saint George St., Toronto, Ontario M5S 1A7, Canada
}%

\author{R. Wagner}
\author{D. Umstadter}
\thanks{current address:  University of Nebraska-Lincoln, Lincoln, NE 68588-0111}
\affiliation{
Center for Ultrafast Optical Science, University of Michigan, Ann Arbor, MI 48109
}%

\author{R.P. Drake}
\affiliation{
Space Physics Research Laboratory, University of Michigan, Ann Arbor, MI 48109-2143
}%

\author{M. C. Downer}
\affiliation{
Department of Physics, The University of Texas at Austin, Austin, Texas 78712
}%

\date{\today}

\begin{abstract}
Detailed spectra of harmonics produced from ultra-intense, sub-picosecond, high-contrast laser pulses incident on solid targets have shown the first observation of regular red- and blue-shifted satellites.  Their frequency shift is slightly less than the frequency of a nominal, pure ion-plasma wave associated with electron critical density, where an ion-acoustic wave would be expected.  We explain this as the result of a substantial reduction of Debye shielding as the intense optical fields compete to drive electrons in large-amplitude oscillations.  This general effect leads to a larger, dynamical and anisotropic Debye length, which should have a broad impact on plasma physics in this regime.
\end{abstract}


\maketitle

\section{Introduction}

Among the phenomena most revealing of new physics in the interaction of ultra-intense coherent light ($I > 10^{18}~\text{W/cm}^{2}$) with matter, harmonic generation at the surface of solids is perhaps the richest.  Distinct from both moderate-intensity surface harmonic generation and from harmonic generation from intense laser pulses in gases, harmonics generated in thin, near-solid-density plasmas, under relativistic interaction conditions, reflect in their detailed spectra the physics of nonlinear scattering from higher-than-rest-mass quivering electrons and from collective electron and ion oscillations, and of hydrodynamics strongly affected by ponderomotive pressure.  Theoretical \cite{Esarey, Wilks, Gibbon, Lichters} and experimental studies \cite{Kohlweyer, vonderLinde, Norreys, Dromey} are well-known to point to qualitative change in the physical relationship of light and matter under these conditions.  New mechanisms have been proposed as nonlinear source terms for optical harmonic generation, among them:  the anharmonic electron current of resonant or of vacuum-heating electrons in the laser field \cite{Gibbon, Brunel}, anharmonic oscillations of the reflective critical-density surface \cite{Lichters, Bulanov}, and the effect of relativistic $J \times B$ coupling \cite{Wilks}.
We report the observation, in detailed harmonic spectra, of multiple, strong, regular, Stokes- and anti-Stokes-like satellite lines.  These reproducible satellites have an abrupt onset intensity, clearly distinct from harmonic production, and appear at the same threshold around each of the several harmonics we studied (3rd through 7th).  Exactly at the appearance of these satellites, the forward-scattered laser-fundamental spectrum shows sharp depletion.  The satellite frequency-shift is the same away from each harmonic, and is the same for red- and blue-shifted peaks in first and second order satellites.  
We propose that these satellites are the result of inelastic scattering from ion waves whose dispersion relation has been modified by the intense laser fundamental pulse.  Near critical density, where the harmonics and probably the satellites are produced, one expects ion-acoustic waves of relatively low frequency.  We describe how large-amplitude electron oscillations can frustrate the Debye shielding of ion charge-fluctuations by electrons, leading to an increased effective Debye length and a premature transition from ion-acoustic to ion-plasma waves.  It is from these modified ion waves that the satellites are produced.
\section{Experimental}
\begin{figure}[ht!]
\includegraphics[width=8.5 cm, trim=0mm 0mm 0 0,clip]{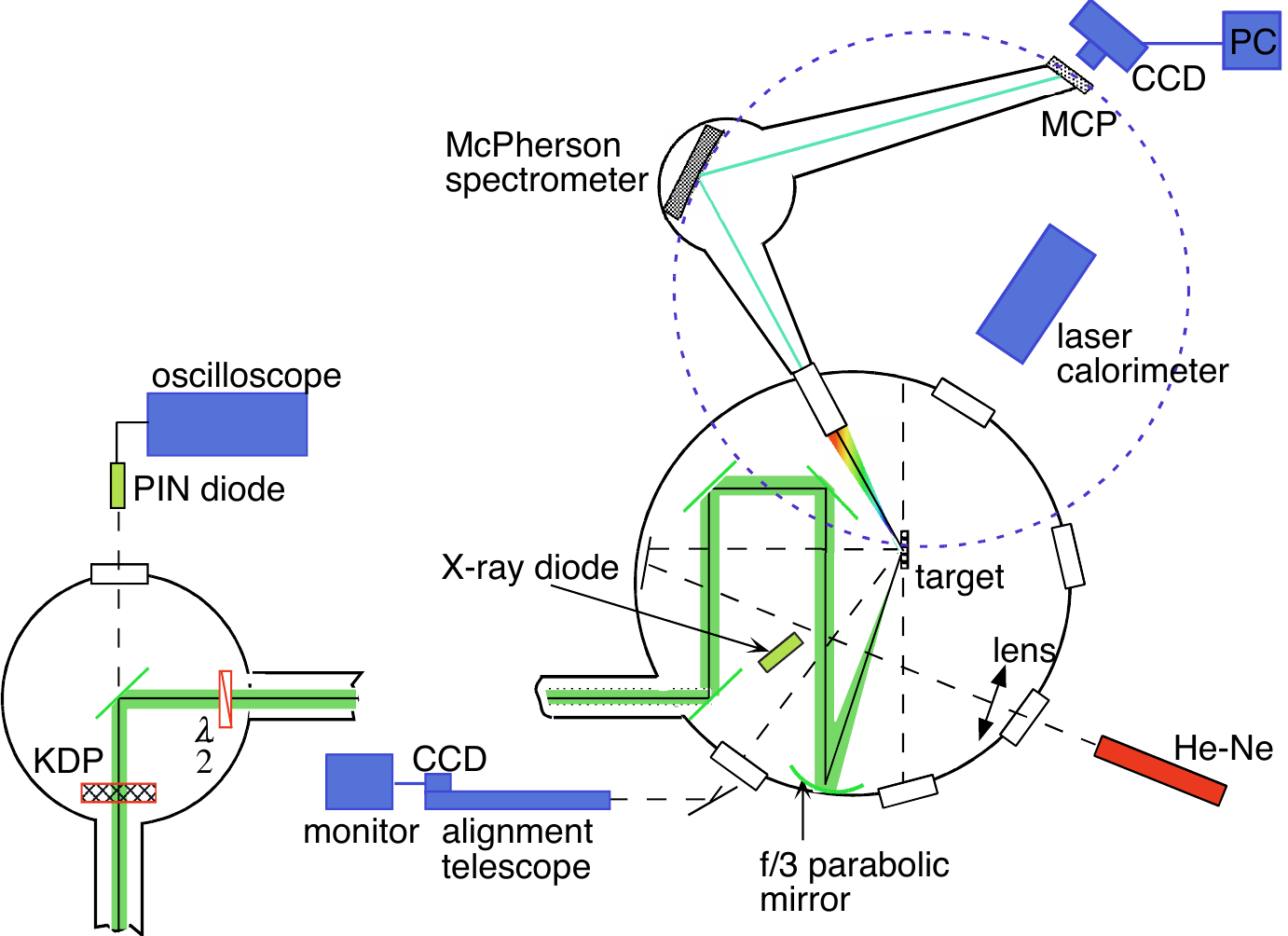}
\caption{\label{fig:Fig1-Layout}  Experimental setup:  The $f=23$cm off-axis parabola focusses the 527-nm laser beam at 60$^{\circ}$ incidence, producing a 9 $\mu$m (FWHM) focal spot.  Specular harmonic spectra are directed into a 1-m VUV spectrometer, using a slit-less configuration, and recorded using a microchannel plate intensifier and 16-bit CCD camera.}
\end{figure}

This work made use of the $T^3$ laser at the Center for Ultrafast Optical Science (CUOS) at the University of Michigan.  This was a hybrid Ti:sapphire and Nd:glass CPA laser system operating at a wavelength of 1.053 µm, producing 400-fs pulses with energies up to 3 J \cite{Mourou}.  Intrinsic pulse contrast is approximately 105;  in order to increase it, the infrared pulse was converted to its second harmonic ($\lambda$ =527 nm) using a 4-mm thick Type-I KD*P crystal \cite{Chien}.  Four harmonic beam-splitters and finally a BG39 glass filter were used after the crystal to filter out the infrared component in the laser beam, producing a green pulse contrast better than $10^{10}$ (Fig. \ref{fig:Fig1-Layout}).

The 43-mm diameter green (527-nm) laser beam was focused at 60$^{\circ}$ incidence onto flat solid targets, at from normal, by an off-axis (15$^{\circ}$) parabolic mirror with 23-cm focal length, producing a 9 $\mu$m (FWHM) focal spot, measured using a partially amplified beam.  Specular harmonic spectra were recorded with a 1-m Seya-Namioka-type VUV spectrometer (McPherson) furnished with a 1200 line mm$^{-1}$ gold-coated grating.  We used a slit-less configuration, in which the tiny laser-plasma source itself was located on the Rowland circle.  At the exit image location on the circle, the detector was a microchannel plate intensifier (Gallileo), its output lens-coupled to a 16-bit CCD camera (Photometrics).  

\begin{figure}
\includegraphics[width=8.5 cm, trim=0mm 0mm 0 0,clip]{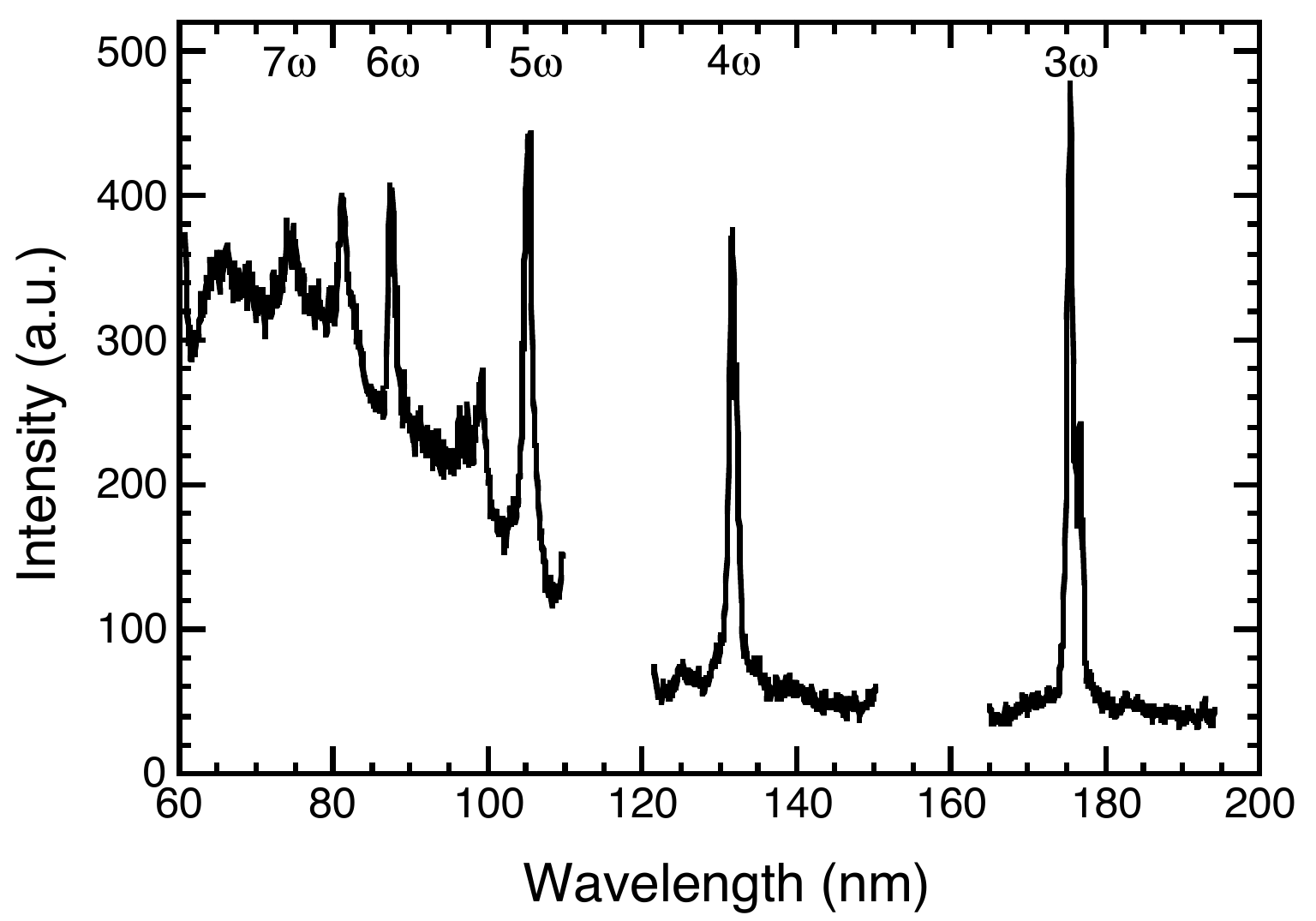}
\caption{\label{fig:Fig2-Spectrum}  A typical time-integrated harmonic spectrum (composite) from Si targets, I = $3.2\times10^{17}~\text{W/cm}^{2}$.}
\end{figure}

Flat targets of different atomic-number materials were used (beryllium, CH, silicon, aluminum, nickel, gold).  In basic respects, the harmonics produced from all of these target materials were similar.  Fig. \ref{fig:Fig2-Spectrum} shows typical time-integrated harmonics from a silicon target irradiated in p-polarization at a laser intensity of $3.2\times10^{17}~\text{W/cm}^{2}$.  Both odd and even orders, from n = 3 to n = 7, appear on top of a broad plasma-recombination background, which background is stronger in the case of high-Z targets.

\section{Observations}

\begin{figure}[ht!]
\includegraphics[width=8.5 cm, trim=0mm 0mm 0 0,clip]{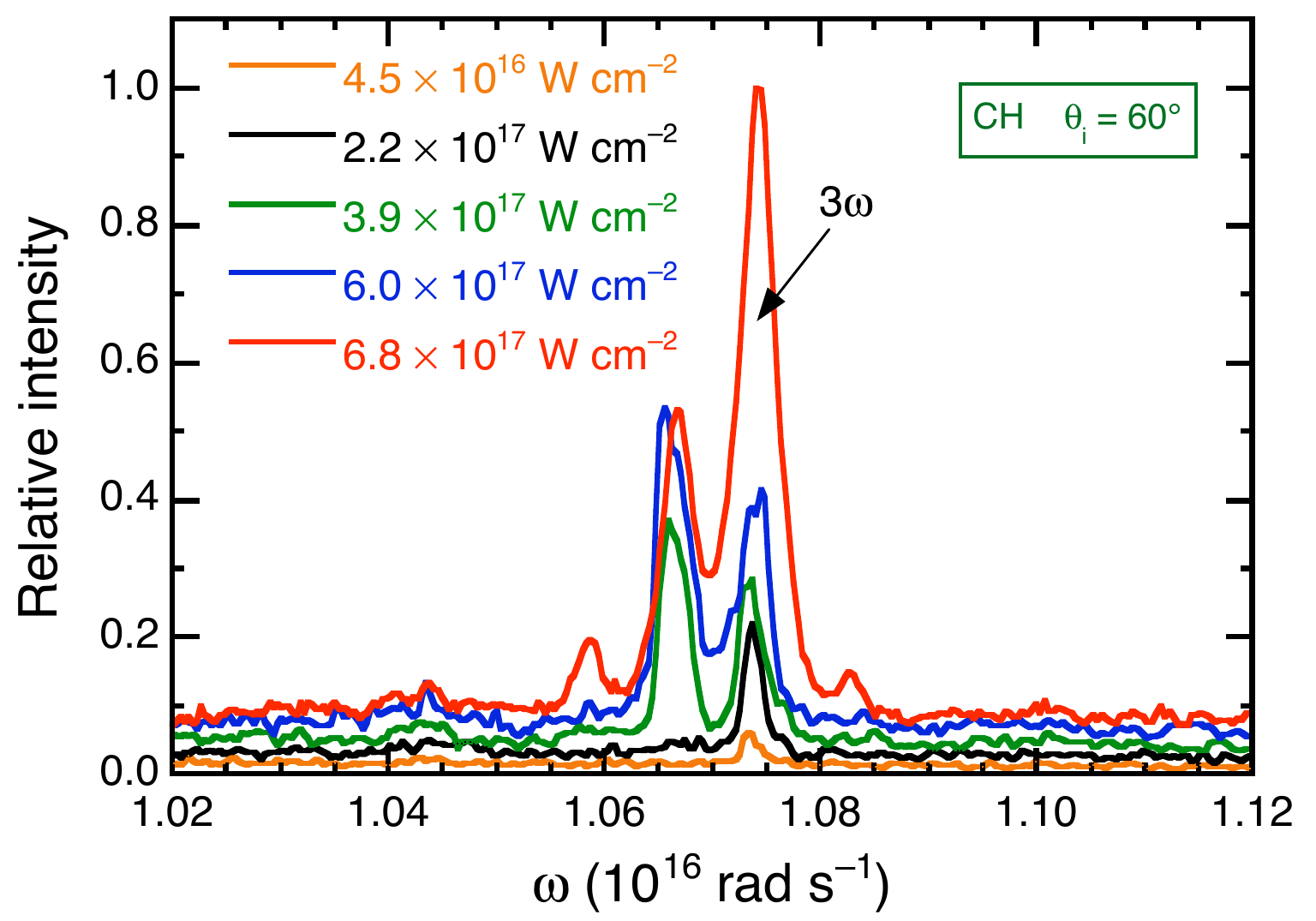}
\caption{\label{fig:Fig3-Satellites}  Detailed harmonic spectra at intensities between $4.5\times10^{16}~\text{W/cm}^{2}$ and $6.8\times10^{17}~\text{W/cm}^{2}$, showing the appearance of regular and reproducible Stokes and anti-Stokes satellites to the harmonics.  Satellites appear on all observed harmonics, and have the same frequency-shift for all harmonics;  the measured shifs suggest a slight dependence on target atomic-number.}
\end{figure}

In resolving the structure of harmonic lines at progressively higher irradiances, we observed \cite{Marjoribanks} the sudden emergence of novel satellite lines, both red- and blue-shifted, having a regular Stokes- and anti-Stokes-like structure. These satellites appeared around each of the 3rd\textendash6th harmonics, apparently simultaneously, but at appreciably higher laser intensities than the appearance of the harmonics themselves \textemdash mid-$10^{17}~\text{W/cm}^{2}$, in our geometry.  As the irradiance was increased to $\sim 1\times10^{18}~\text{W/cm}^{2}$, we observed the sequential appearance of three such peaks:  first a red-shifted peak, then a blue-shifted peak, then an additional red-shifted peak, each stepped in frequency by the same increment (Fig.  \ref{fig:Fig3-Satellites}).  The satellites were repeatable and spectrally narrow.  In a few cases, the line on the longer-wavelength side was as intense as the harmonic line itself.

\begin{figure}[ht!]
\includegraphics[width=8.5 cm, trim=0mm 0mm 0 0,clip]{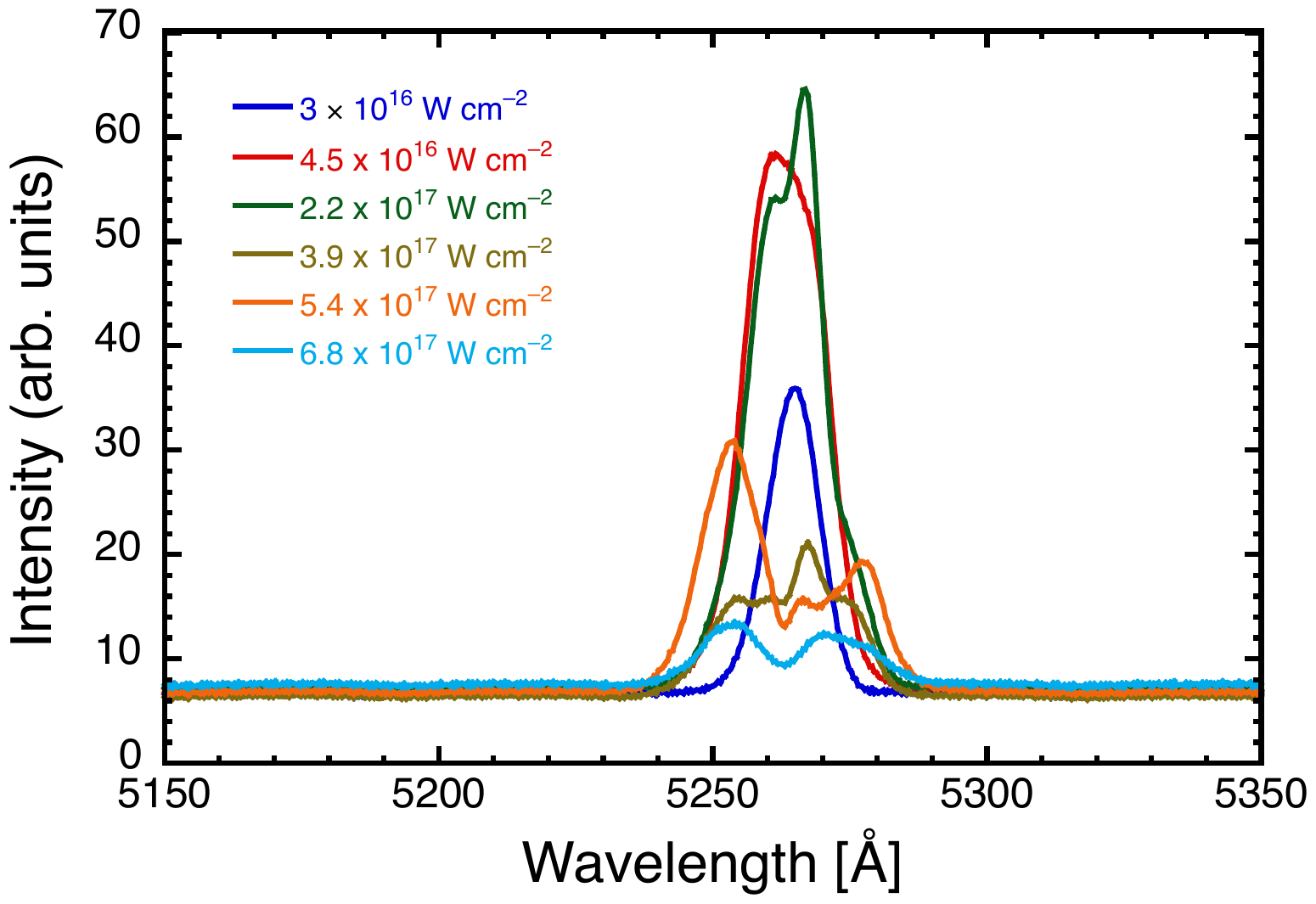}
\caption{\label{fig:Fig4-ForwardScatt}  Detailed spectra of the specular-scattered laser fundamental, at intensities between $3\times10^{16}~\text{W/cm}^{2}$ and $1\times10^{18}~\text{W/cm}^{2}$, showing the sudden depletion of pump-laser reflected light coinciding precisely with the appearance of satellites.}
\end{figure}

These lines appeared for different atomic-number materials:  Be, CH, Al, and Si, but were not seen under any of our conditions for the higher-Z elements, Ni and Au.  For CH targets, the frequency step between satellites was $\sim7.4\times10^{13}~\text{rad/s}$.  The data are suggestive that this shift may be weakly Z-dependent, with a 10\% difference between Be and Si.
At our detection threshold of $10^{-4}$ of incident intensity, we detected no backscattered light.  However, the spectrum of forward-scattered laser light showed sudden line-centre depletion exactly upon the appearance of the satellite features (Fig.  \ref{fig:Fig4-ForwardScatt});  the spectrum appears broader but in fact this is in the context of the reduced peak \textemdash the wings of the laser spectrum are approximately unaffected.  

The sudden depletion of the forward-scattered pump light, coinciding with the appearance of these regular satellites, argues strongly for an onset threshold for driving up new waves in the plasma, from which the pre-existing harmonics inelastically scatter to produce satellites.  The harmonics themselves are widely considered \cite{harmonicsCrit} to originate at the critical density surface, where electron plasma waves are strongly enhanced and where anharmonic oscillations of the critical surface yield higher-harmonic Fourier content.

\section{Discussion}

Since mid- and high-harmonics are widely accepted to originate at the critical density surface, and since the satellites appear in all harmonics together with the same frequency shift, it seems clear that the satellites also originate at critical during the generation of harmonics \textemdash ~these are not satellites associated with the fundamental, since they would subsequently be converted together with the harmonics into different orders.  There are few candidate plasma waves, associated with the critical density surface and with frequency near the observed satellite shift.  The most immediate prospect, and the only one we have identified in this context, is the oscillation of the ions near critical.  Pure ion-plasma waves associated with near-fully ionized plasma around critical would have a frequency about 60\% of the observed shift.  This suggests one possible prospect:  strong driven oscillations of electrons at critical may reach a threshold to drive up an ion-wave structure there;  the electromagnetic wave which drives the electron oscillations and which scatters to produce harmonics, now scatters inelastically from this ion structure.  Such a driven ion wave spectrum will comprise k-vectors into and out of the plasma.  In this picture the electromagnetic wave can then downscatter in energy, to excite two opposing wavevectors, or upscatter, taking energy from two opposing wavevectors.  In either case, the k-vector diagram of the scattered light remains specular for the satellites, as for the harmonics.
[[may need reference to width of peak, from earlier paper...]]
The description begs the question that ion-plasma frequencies are accessible at all, though simple calculations indicate oscillations should be ion-acoustic at this point.  Heuristically, ion density perturbations should be dressed by the plasma electrons Debye-shielding them, wherein the restoring force is set by electron plasma pressure.  Ion plasma waves result for $k\lambda D >> 1$, in which case the scale of Debye shielding is larger than the wavelength of the ion wave, and is ineffective in screening ion charge-excesses;  in that case, the ion perturbations see a much stronger restoring electrostatic force due to the ion perturbations themselves, and the much-higher ion-plasma frequency results.
In that case, the argument reduces to understanding how Debye shielding can be reduced, near critical density in intense laser-matter interaction.  This may come about if the assumption of quasineutrality is very substantially violated near critical in a steep density gradient \textemdash possibly the electrons are substantially pushed off the ions there, and cannot participate in Debye shielding.  It may be the case also that sufficiently steep density gradients may alter ion waves, should it be that $\lambda\pi > L$.  The most plausible possibility seems, though, to be the observation that electrons driven in a strong laser field begin to lose their ability to Debye shield ion-wave perturbations, once their oscillation amplitude approaches half the wavelength of an ion wave.

An analytic description of this phenomenon, in a nonrelativistic, homogeneous fluid model, has already been published \cite{Drake}.  There is also a simple kinematic picture which may be more intuitive:  Debye shielding describes an electrostatic potential which is self-consistent with the kinetic energy distribution of electrons, or, in force terms, the Coulomb attraction of electrons for a test charge, balanced against the electron pressure.  In this context the temperature is the most important moment of the electron velocity distribution.  In a high-intensity laser-plasma, the electron quiver motion gives an additional contribution to the velocity distribution, and provides an anisotropic contribution to the pressure through its average kinetic energy, described through its second moment as an effective temperature.
\label{eq:LambdaDebye}
\begin{eqnarray}
\lambda_D \equiv (\frac{\epsilon_0 k_B T_e}{n e^2})^{1/2}
\end{eqnarray}
Heuristically, we can capture much of the most relevant physics of Debye shielding in an intense laser field by adding to the symmetric thermal kinetic energy the orientation-dependent kinetic energy of oscillation.  The temperature, the second moment of the distribution, still represents the average kinetic energy, provided that the temperature $T_e$ is replaced by an anisotropic $T_e^{\star}$, just as:
\label{eq:VTherm}
\begin{eqnarray}
v_{th}^2 \rightarrow v_{th}^2 + (\frac{k_z}{k})^2 v_{os}^2
\end{eqnarray}
Here the direction cosine ($k_z/k$) reproduces the contribution of the electron oscillation in any given direction.  Thus we substitute:
\label{eq:Taniso}
\begin{eqnarray}
T_e  \rightarrow T_e^{\star} \equiv T_e (1+(\frac{k_z}{k})^2 (\frac{v_{os}}{v_{th}})^2)
\end{eqnarray}
In the standard formula for the Debye length, this anisotropic temperature leads to an increased Debye length, according to:
\label{eq:LambdaDebyeAniso}
\begin{eqnarray}
\lambda_D^{\star} = (\frac{\epsilon_0 k_B T_e^{\star}}{n e^2})^{1/2}
\end{eqnarray}
thus
\label{eq:Remapk2L2}
\begin{eqnarray}
k^2 \lambda_D^2  \rightarrow k^2 {\lambda_D^{\star}}^2 = k^2 \lambda_D^2  + k_z^2 \lambda_S^2 
\end{eqnarray}
where
\label{eq:LambdaS}
\begin{eqnarray}
\lambda_S  \equiv  \frac{v_{os}}{\sqrt{2} \omega_{p,e}}
\end{eqnarray}
Thus the Debye sphere is stretched, only along the direction of electron oscillation, to become a Debye prolate-ellipsoid.  In light of this, the conventional dispersion relation for ion waves then is modified in a straightforward way:
\label{eq:NewDisp}
\begin{eqnarray}
\omega^2 - 3 k^2 \frac{T_i}{M} = \frac{\omega_{p,i}^2 (k^2 \lambda_D^2  + k_z^2 \lambda_S^2)}{(1+k^2 \lambda_D^2  + k_z^2 \lambda_S^2)}
\end{eqnarray}
Therefore, in a strong oscillating electric field, the frequency of oscillation of an ion wave in a quasineutral plasma can be greatly increased, as Debye shielding is partially suppressed, and electrostatic restoring forces are free to play an increasing role.  

It's illustrative to roughly sketch conditions for frustrating Debye shielding, around the critical surface, sufficient to shift ion waves to a frequency 0.6 $\omega_{p,i}$, as inferred here.  Starting from cold ions and an assumption that in the absence of the laser field Debye shielding of relevant ion density perturbations is effective ($k\lambda_D <<$ 1 [otherwise more heavily damped]), we find we require $k_z^2 \lambda_S^2 = k^2 cos(\theta)^2 \lambda_S^2$ = 0.56, where $\theta = 60^\circ$.  Assuming gross motion, and so ion waves of wavelength comparable to the density scalelength $L\sim \lambda / 10$ and longer, we the find a condition that $a_0 = v_{os}/c > 0.2$, which at our wavelength implies an onset intensity greater than a few $10^{17}~\text{W/cm}^{2}$, comparable to observation.  More simply, this condition is roughly equivalent to requiring that $x_{os}$, the quiver distance, be comparable to the plasma gradient scalelength $L$.  These are qualitative estimates, of course, and kinetic effects and the impact of inhomogeneity will be significant.


In conclusion, we have reported new observations of regular Stokes-like and anti-Stokes-like satellite features accompanying the harmonics.  We speculate that these result from inelastic scattering from ion-waves driven up by strong electron oscillations.  The frequency of such waves is much nearer to ion-plasma waves than to ion-acoustic waves one would likely expect, and we explain this in terms of modification of the ion-wave dispersion relation, caused by the frustration of Debye shielding by electrons subject to strong oscillating fields.

\section{Acknowledgements}

This work was supported by the Natural Sciences and Engineering Research Council of Canada, Photonics Research Ontario, the Fellows Program of the Center for Ultrafast Optical Science at the University of Michigan, and by the National Science Foundation through its support of its Institutes of Science and Technology. DPU: grants NSF PHY 972661 and NSF STC PHY 8920108;  SPL, MCD: partial support of DOE grant DE-FG03-97ER54439, Robert Welch Foundation grant F-1038, and a Faculty Research Assignment from the University of Texas.

\end{document}